\documentclass[sigconf]{acmart}

\usepackage{graphicx}
\usepackage{tabularx}
\usepackage{booktabs}
\usepackage{array}
\usepackage{tabularx}
\usepackage{hyphenat}

\AtBeginDocument{%
  \providecommand\BibTeX{{%
    \normalfont B\kern-0.5em{\scshape i\kern-0.25em b}\kern-0.8em\TeX}}}

\copyrightyear{2026}
\acmYear{2026}
\setcopyright{cc}
\setcctype{by}
\acmConference[CSCW Companion '26]{Companion of the Computer-Supported Cooperative Work and Social Computing}{October 10--14, 2026}{Salt Lake City, UT, USA}
\acmBooktitle{Companion of the Computer-Supported Cooperative Work and Social Computing (CSCW Companion '26), October 10--14, 2026, Salt Lake City, UT, USA}
\acmDOI{10.1145/3785651.3818822}
\acmISBN{979-8-4007-2378-0/2026/10}

\begin{document}

\title[Contribution, Originality, and Accountability with Agents in Collaboration]{"Nobody Did This": Contribution, Originality, and Accountability in Agent-Mediated Collaboration}

\author{Kashif Imteyaz}
\affiliation{%
  \institution{Northeastern University}
  \city{Boston}
  \country{USA}}
\email{imteyaz.k@northeastern.edu}

\author{Mohammad Rashidujjaman Rifat}
\affiliation{%
  \institution{University of Notre Dame}
  \city{Notre Dame}
  \country{USA}}
\email{rifat@nd.edu}

\author{Divya Ramesh}
\affiliation{%
  \institution{University of North Carolina at Charlotte \& Mohamed bin Zayed University of Artificial Intelligence (MBZUAI)}
  \city{Charlotte}
  \country{USA}}
\email{dramesh4@charlotte.edu}

\author{Steven R. Rick}
\affiliation{%
  \institution{NVIDIA}
  \city{Santa Clara, CA}
  \country{USA}}
\email{srick@nvidia.com}

\author{Simo Hosio}
\affiliation{%
  \institution{University of Oulu}
  \country{Finland}}
\email{simo.hosio@oulu.fi}

\author{Hauke Sandhaus}
\affiliation{%
  \institution{Cornell Tech}
  \city{New York}
  \country{USA}}
\email{hgs52@cornell.edu}

\author{Advait Sarkar}
\email{advait@microsoft.com}
\affiliation{%
 \institution{Microsoft Research \& University of Cambridge \& University College London}
 \city{Cambridge}
  \country{United Kingdom}
}

\author{Christoph Riedl}
\affiliation{%
  \institution{Northeastern University}
  \city{Boston}
  \country{USA}}
\email{c.riedl@northeastern.edu}

\author{Saiph Savage}
\email{s.savage@northeastern.edu}
\affiliation{%
  \institution{Civic A.I. Lab, Northeastern University}
  \city{Boston}
  \state{Massachusetts}
  \country{USA}
}
\affiliation{%
  \institution{Universidad Nacional Autonoma de Mexico (UNAM)}
  \city{Mexico City}
  \country{Mexico}
}

\renewcommand{\shortauthors}{Imteyaz et al.}

\begin{abstract}
Collaborative knowledge work is changing in ways that go beyond disclosure or transparency. LLM agents are now embedded in how teams research, design, write, and decide: mediating between members, synthesizing inputs, reformulating ideas, and drafting shared outputs. They do not only facilitate collaboration; they operate within the workflow at the moment contributions are being formed. In doing so, they risk undermining the social conditions under which contribution can be witnessed, attributed, and held accountable. This workshop brings together researchers and practitioners to confront what we call contribution dissolution: the blurring of attribution, originality, and accountability in agent-mediated collaborative work. We argue that this dissolution begins before collaboration itself, in the individual worker's own uncertainty about what is genuinely theirs, and propagates through collaborative relationships, collapsing the reliability that makes productive intellectual exchange possible. Through position statements, mapping exercises, and a hands-on activity, participants will surface how framing accountability as a documentation problem (e.g., AI use statements, watermarking, provenance logs) overlooks the conditions under which accountability is produced. Our goal is to produce a shared research agenda and the foundations of an infrastructural response to contribution dissolution in collaborative knowledge work.
\end{abstract}


\maketitle

\section{Introduction}
Collaborative knowledge work rests on an implicit social contract: people who work together maintain a shared, informal record of who contributed what. This record lives in memory, in conversation traces, in the history of drafts and revisions \cite{birnholtz2012tracking}. Drawing on Suchman's account of accountable action as situationally embedded \cite{suchman1987plans}, we call this \textit{witnessed contribution}: a contribution that can be attributed, contested, and accounted for because it emerged within a shared social situation that made it intelligible to others. Witnessed contribution is not the same as documented provenance. Provenance records that something was produced and how it was produced. What makes a contribution attributable and accountable is that it was produced under conditions where others could recognize it as an act of judgment, respond to it, and hold the contributor(s) answerable. A provenance log cannot show whether a framing was the person's or the agent's, or whether the person could defend it under challenge. These require the kind of situated social interaction that agent mediation risks undermining.

LLM agents shift this at the individual level first \cite{he2025exploring}. Rather than using agents to externalize already-formed ideas, people co-think with them, with ideas emerging through interaction \cite{hoque2024hallmark, heyman2024supermind}. They draft, refine, and develop lines of reasoning in dialogue with these agents \cite{han2024teams}. As a result, the boundary between what the individual contributed and what the agent introduced becomes difficult to reconstruct after the fact \cite{zindulka2025ai, he2025exploring}. It was never established during the process itself. A worker may have procedural ownership (they prompted, reviewed, and accepted) without having substantive authorship that could survive scrutiny. For instance, Zindulka et al. \cite{zindulka2025ai} show that workers often fail to remember the source of content created in human-AI workflows, with accuracy dropping most in mixed workflows where some steps were human and others agent-mediated. The individual worker arrives at the collaborative table already unable to reliably say what is theirs.

At the collaborative level, agents intervene at the moment contributions are formed and shared \cite{mysore2025prototypical}. When a team member uses an agent to reformulate an idea before sharing it \cite{ulloa2025product, he2024ai}, when an agent reconciles two collaborators' positions into a synthesized output, when an agent drafts a deliverable from multiple inputs, the contribution that reaches collaborators is not what the person said or wrote but what the agent produced from what they said or wrote \cite{draxler2024ai}. Social creativity depends on stakeholders externalizing genuinely distinct perspectives; it is the friction between different framings that generates new understanding \cite{fischer1999symmetry}. Agents are optimized to follow instructions, not to resist them. A human collaborator pushes back, asks why, offers a competing frame. An agent takes what it is given and produces the most plausible completion, shaped by the conventions and biases embedded in its training data. The result is convergence: outputs that reflect dominant assumptions rather than the situated insight of the person prompting \cite{haraway2013situated}. What is lost is not just stylistic distinctiveness but the particular knowledge a person brings from their practice, their domain, the specific conditions under which they encountered the problem. Agent refinement replaces this with a generic framing while appearing to have improved only the surface.

When a colleague shares agent-mediated work, the recipient cannot read it as a reliable signal of that colleague's thinking \cite{imteyaz2026co, he2025contributions}. They do not know whether the contributor can defend it, whether their response to pushback represents a real intellectual position or a new prompt, or whether the voice in the document belongs to the person or the model. The contributor themselves may not know \cite{zindulka2025ai}. When neither the person sharing work nor the person receiving it can distinguish human contribution from agent reformulation \cite{he2025contributions}, the shared understanding that makes collaboration productive breaks down. Work continues to circulate, but it no longer reliably reflects the exclusive thinking of the people producing it. We call this \textit{contribution dissolution}: not a failure of disclosure but a condition in which the social basis for attribution, originality, and accountability has been blurred.

Common responses in the research community (disclosure-based solutions, AI use statements, watermarking, provenance tracking) frame contribution dissolution as a problem of attribution and documentation: if we could better track what the agent did, we could restore accountability \cite{birnholtz2012tracking, liu2024survey, weisz2025human, he2025exploring}. This framing is productive in cases where human and agent contributions remain separable, for instance when an agent is used to check grammar or reformat a document. However, when a team member uses an agent to "clean up" collaborative notes before sharing them, the cleanup may restructure the argument in ways that redistribute credit for key ideas. Watermarks may not persist through ordinary collaborative refinement. Detection tools fail in tightly coupled human-agent work where outputs converge. Disclosure cannot surface what was never witnessed, or what, as Zindulka et al. \cite{zindulka2025ai} show, was never reliably encoded in the contributor's own memory. The issue is not that documentation-based approaches are misguided, but that contribution dissolution may involve dimensions that go beyond what current documentation-based approaches were designed to capture.

This workshop is situated at the intersection of several active CSCW research threads (accountability in cooperative work, the social organization of knowledge production, and human-AI collaboration) and brings them into productive tension around a problem none has fully confronted. We build on foundational CSCW work on visible and invisible work \cite{star1999}, accountability in situated action and human-machine reconfigurations \cite{suchman1987plans, suchman2007human}, and recent work on AI failure loops in workplace contexts \cite{kawakami2026ai, jo2025understanding} to argue that contribution dissolution is not a peripheral concern but a structural transformation of what collaborative knowledge work means.

\subsection{Workshop Themes}
The workshop is organized around three themes:

\begin{itemize} \item \textbf{Contribution Dissolution:} How do agents restructure the process through which contributions are formed, and what is lost when they intervene inside the collaborative loop? This theme addresses both the individual level (the worker's uncertainty about what is genuinely theirs \cite{zindulka2025ai, pachera2026co}) and the collaborative level (the collapse of productive asymmetry between contributors when agent refinement makes distinct perspectives indistinguishable \cite{fischer1999symmetry, imteyaz2026co}).

\item \textbf{The Documentation Trap:} How do documentation-based responses (AI use statements, detection tools, watermarking, provenance tracking) fail, and what does this failure reveal? Our central argument is that these solutions approach contribution dissolution as a problem of missing information. They assume that the evidence needed for attribution and accountability exists and merely needs to be surfaced. This theme confronts the possibility that the information was never produced, and asks what follows for the governance of AI in collaborative work. 

\item \textbf{Accountability Infrastructures:} How should we think about accountability relations given agent-mediated work? What might the composition of appropriate forums look like to judge agent-mediated work? If documentation can no longer be reliable sources of attribution due to blurred human-agent boundaries, what would accountability infrastructures require? This theme moves from diagnosis to response, asking what collaborative systems, institutional norms, and governance mechanisms would need to preserve for witnessed contribution to remain possible in agent-mediated work.

\end{itemize}

\section{Workshop Activities}

\subsection{Main Workshop Schedule}
Table \ref{tab:schedule} presents the interactive session format, with the planned activities organized into two phases. The workshop requires a projector, whiteboard or large poster paper, sticky notes, and markers.

\begin{table}[!htbp]
\footnotesize
\centering

\label{tab:schedule}
\renewcommand{\arraystretch}{1.1}

\begin{tabular}{p{0.14\columnwidth}
                p{0.22\columnwidth}
                p{0.56\columnwidth}}
\toprule

\textbf{Time} &
\textbf{Activity} &
\textbf{Description} \\
\midrule

09:00--09:30 &
Opening and Framing &
Introduction to the workshop’s central provocation and participant introductions. \\

09:30--10:15 &
Lightning Talks &
Short provocations on attribution, originality, and accountability under agent mediation. \\

10:15--10:30 &
Coffee Break &
\\

10:30--11:15 &
Mapping Session I &
Groups identify moments where contribution disappears during agent-mediated collaboration. \\

11:15--12:00 &
Mapping Session II &
Groups reconstruct attribution from logs and outputs, surfacing the limits of provenance mechanisms. \\

12:00--13:00 &
Lunch &
\\

13:00--13:30 &
Reframing the Problem &
Whole-group synthesis and discussion of alternatives beyond documentation. \\

13:30--14:30 &
Design Futures I &
Groups brainstorm accountability infrastructures across collaborative domains. \\

14:30--14:45 &
Coffee Break &
\\

14:45--15:30 &
Design Futures II &
Cross-context synthesis and drafting contributions to the casebook. \\

15:30--16:00 &
Research Agenda Building &
Participants identify empirical questions and future research directions. \\

16:00--16:30 &
Synthesis and Next Steps &
Presentation of the draft casebook and discussion of future collaborations. \\

\bottomrule
\end{tabular}
\caption{Proposed full-day workshop schedule.}
\label{tab:schedule}
\end{table}

\subsection{The Documentation Trap Activity}
Participants receive a package of materials representing a completed collaborative project: a final report, a set of intermediate drafts, and a provenance log recording all agent interactions throughout. Groups are asked to answer four questions from the materials alone: 
\begin{itemize}
    \item Who had the original idea for the central framing?
    \item Who had the original idea for the central framing?
   \item Whose position won when two collaborators disagreed? 
   \item Which contributions survived agent reformulation, and which were lost?
   
   \item Who should be held accountable if the report's central claim turns out to be wrong?

\end{itemize}
Groups will discover that the provenance record offers no definitive answers to these questions. The debrief discussion surfaces why: this is not a failure of the provenance log. It is a limitation of what the log was asked to record.

The materials will be constructed by the organizers to represent a realistic collaborative workflow involving multiple contributors and agent interactions at different stages. Groups of 4--5 participants will have 30 minutes to work through the questions, annotating the provenance log to mark where they can and cannot establish attribution. Each group will then briefly report their findings, and the facilitator will lead a debrief discussion comparing what the logs could and could not reveal. This activity feeds directly into the afternoon session on accountability infrastructure, where participants draw on their experience of the activity's limitations to inform design alternatives.

\section{Organizers}





We introduce the organizing team, consisting of junior and senior researchers from academia and industry who bring diverse expertise and complementary perspectives to the workshop.

\textbf{Kashif Imteyaz} (lead Facilitator \& Proceedings Chair) is a PhD Candidate at Northeastern University. His work explores the design and evaluation of generative AI systems that support how people think, work together, and co-create with AI \cite{imteyaz2026co, imteyaz2026upskilling}. He also chairs the ACM SIGCHI Boston chapter.

\textbf{Mohammad Rashidujjaman Rifat} (Proceedings Chair) is an Assistant Professor of Technology Ethics and Global Affairs at the University of Notre Dame. His research sits at the intersection of HCI, ethics in AI, and critical social science, and focuses on how computing systems can support coexistence in plural societies~\cite{rifat2023many, rifat2024cohabitant, rifat2024data}.

\textbf{Divya Ramesh} (Proceedings Chair) is an Assistant Professor of Human-Centered Computing at the University of North Carolina at Charlotte, USA, and a visiting scholar of Human-Computer Interaction at Mohamed bin Zayed University of AI (MBZUAI), UAE. She leads the Bridges for Responsible Computing group that combines interdisciplinary perspectives to advance accountable AI innovation for economic inclusion~\cite{ramesh2022platform, ramesh2023ludification}. 

\textbf{Steven R. Rick} (Publicity Chair) is a Solutions Architect at NVIDIA. He received his Ph.D. in Computer Science from the University of California, San Diego. Steven focuses on how to design and develop systems that combine humans and computers to leverage the strengths of each, and co-led a 2025 CSCW workshop on GenAI's role in collaborative group-work\cite{johnson2025augmenting}.

\textbf{Simo Hosio} is a professor of Computer Science and Engineering at the University of Oulu, Finland. Simo leads the Crowd Computing research group and has organised several workshops across key HCI venues (CSCW, CHI, UbiComp). His work focuses on applied AI in digital health, human-AI interaction, and, relevant to this proposal, human-centric data management (e.g. data valuation, monetary or otherwise).  

\textbf{Hauke Sandhaus} (Web and Publicity Chair) is a PhD candidate at Cornell Tech. His research examines how generative AI shapes interactive system design, how students navigate self-initiated AI use in HCI workflows, and issues of transparency in user research. He also teaches workshops and summer courses on design ethics.

\textbf{Advait Sarkar} is a Senior Researcher at Microsoft Research in Cambridge and an affiliated lecturer at University of Cambridge, as well as an honorary lecturer at University College London. His research explores human-centred AI, end-user programming, and how people interact with data and intelligent systems.

\textbf{Christoph Riedl} is a professor for Information Systems and Network Science at the D’Amore-McKim School of Business at Northeastern University. He is the chair of the ACM Collective Intelligence Conference Steering Committee, a fellow at the Institute for Quantitative Social Science (IQSS) at Harvard, and a fellow at the Center for Collective Intelligence at MIT.

\textbf{Saiph Savage} is an Assistant Professor and Director of the Civic AI Lab at Northeastern University, where she co-designs, develops, and studies public AI technologies that empower workers \cite{savagegigsense, savage2026ai}, federal agencies, industry leaders, and NGOs \cite{saiph2024human}.

\subsection{Before the Workshop}
Our goal is to bring together a multidisciplinary group of researchers and practitioners from HCI, CSCW, STS, labor studies, AI ethics, and design who engage with the problem of contribution dissolution. The organizing team reflects this range and will recruit through ACM SIGCHI and CSCW mailing lists, the CHIWORK community, FAccT networks, and direct outreach to researchers studying human-AI collaboration, cooperative work, and labor. To ensure practitioner representation, we will also recruit through industry channels, including UX and design professional communities, software engineering forums, and the organizers' professional networks in industry research labs. We aim to involve about 25–30 participants, balancing academic and practitioner perspectives to ground the workshop's discussions in both theoretical and professional experience.

To curate an engaged program, we will form a small Program Committee. Each submission will be reviewed by at least two PC members, focusing on the concreteness of the case, engagement with workshop themes, and potential for discussion. We will prioritize submissions that present empirical encounters with contribution dissolution from practice, research, or professional experience rather than purely theoretical work. Selection will also consider diversity in career stage, discipline, and practitioner versus academic perspective.

Accepted participants will be invited to a dedicated Slack channel before the workshop, where they will share brief introductions and a short summary of their position provocation. This asynchronous exchange allows participants to familiarize themselves with each other's cases and perspectives without adding significant preparation burden. Organizers will cluster the submissions thematically in advance to guide the mapping activity and ground discussions in concrete cases from the start.

\subsection{Post-Workshop Plans \& Dissemination}
All accepted materials (position provocations, slides, and artifacts generated during the workshop) will be made available on the workshop website under a CC BY license. The primary output will be a collaboratively developed \textbf{Accountability Infrastructure Casebook}: a design provocation, grounded in the concrete cases participants bring, outlining what collaborative systems and institutional norms would need to be preserved for witnessed contribution to remain possible. This casebook will be drawn from the position provocations and refined through the mapping and brainstorming activities during the workshop.

To reach a wider audience, we will archive accepted position provocations and workshop artifacts with DOIs on Zenodo under an open license. In the long term, we will synthesize the findings into a collaboratively authored paper for CSCW or CHI and disseminate a summary in ACM Interactions. Participants will be invited to continue through a post-workshop working group focused on the governance implications of contribution dissolution across professional and institutional settings.

\section{Accessibility}
Accessibility will be integrated throughout the workshop. Authors will be required to follow the SIGCHI Accessible Submission Guide\footnote{Guide to an Accessible Submission: \url{https://sigchi.org/conferences/author-resources/accessibility-guide/}}. We will review submissions for accessibility compliance, collect information about participant accessibility needs in advance, and coordinate with the CSCW Accessibility Chairs during the event. All workshop materials will include alt text and other accessible features.

\section{Call for Participation}
This workshop confronts contribution dissolution: the condition in which agent-mediated collaborative work blurs the social basis for attribution, originality, and accountability. We invite 2--4 page position provocations from researchers, practitioners, and theorists working on human-AI collaboration, accountability in cooperative work, or the social organization of knowledge production. Position provocations should take a clear stance on one or more of the following themes:
\begin{itemize}
\item \textbf{Contribution Dissolution:} How do agents restructure the process through which contributions are formed, and what is lost when they intervene inside the collaborative loop? We welcome cases from any professional domain where the boundary between human and agent contribution became difficult or impossible to establish.
\item \textbf{The Documentation Trap:} Why do documentation-based responses (AI use statements, detection tools, watermarking, provenance tracking) fall short, and what does this reveal? We are interested in cases where provenance records existed but could not answer the questions that mattered.
\item \textbf{Accountability Infrastructure:} If documentation cannot recover what was lost, what would accountability infrastructure require? We welcome design provocations, governance proposals, or institutional experiments that move beyond disclosure toward new forms of collaborative accountability.
\end{itemize}
We particularly welcome provocations grounded in concrete cases of contribution dissolution from professional settings including research, design, software engineering, and content creation.
\begin{itemize}
\item Submissions should be 2--4 pages excluding references, following ACM CSCW formatting guidelines, and submitted via Google Form.
\item We will offer two submission deadlines: an early deadline (July 10, 2026) to support participants who require advance planning for travel or visa applications, and a later deadline (August 1, 2026) for those already planning to attend the conference.
\item Selection will be based on the concreteness of the case or argument and the diversity of perspectives represented.
\item At least one author of each accepted submission must attend the workshop.
\end{itemize}

\begin{acks}
The authors assert full authorship over this proposal. Planning steps and thematic decisions were made entirely without AI assistance. We used Claude and Grammarly as writing assistants to refine language, streamline the structure, and ensure clarity, retaining complete oversight.

\end{acks}

\bibliographystyle{ACM-Reference-Format}
\bibliography{workshop-proposal}

@book{suchman2007human,
  title={Human-machine reconfigurations: Plans and situated actions},
  author={Suchman, Lucille Alice},
  year={2007},
  publisher={Cambridge university press}
}

@book{suchman1987plans,
  title={Plans and situated actions: The problem of human-machine communication},
  author={Suchman, Lucille Alice},
  year={1987},
  publisher={Cambridge university press}
}

@incollection{haraway2013situated,
  title={Situated knowledges: The science question in feminism and the privilege of partial perspective 1},
  author={Haraway, Donna},
  booktitle={Women, science, and technology},
  pages={455--472},
  year={2013},
  publisher={Routledge}
}

@inproceedings{fischer1999symmetry,
  title={Symmetry of igorance, social creativity, and meta-design},
  author={Fischer, Gerhard},
  booktitle={Proceedings of the 3rd Conference on Creativity \& Cognition},
  pages={116--123},
  year={1999}
}

@article{star1999,
  title={Layers of silence, arenas of voice: The ecology of visible and invisible work},
  author={Star, Susan Leigh and Strauss, Anselm},
  journal={Computer supported cooperative work (CSCW)},
  volume={8},
  number={1},
  pages={9--30},
  year={1999},
  publisher={Springer}
}

@article{imteyaz2026co,
  title={Co-Designing Collaborative Generative AI Tools for Freelancers},
  author={Imteyaz, Kashif and Muller, Michael and Flores-Saviaga, Claudia and Savage, Saiph},
  journal={arXiv preprint arXiv:2602.05299},
  year={2026}
}

@article{kawakami2026ai,
  title={AI failure loops in devalued work: The confluence of overconfidence in AI and underconfidence in worker expertise},
  author={Kawakami, Anna and Taylor, Jordan and Fox, Sarah and Zhu, Haiyi and Holstein, Kenneth},
  journal={Big Data \& Society},
  volume={13},
  number={1},
  pages={20539517261424164},
  year={2026},
  publisher={SAGE Publications Sage UK: London, England}
}

@inproceedings{jo2025understanding,
  title={Understanding Public Agencies' Expectations and Realities of AI-Driven Chatbots for Public Health Monitoring},
  author={Jo, Eunkyung and Kim, Young-Ho and Ok, Sang-Houn and Epstein, Daniel A},
  booktitle={Proceedings of the 2025 CHI Conference on Human Factors in Computing Systems},
  pages={1--17},
  year={2025}
}

@article{zindulka2025ai,
  title={The AI Memory Gap: Users Misremember What They Created With AI or Without},
  author={Zindulka, Tim and Goller, Sven and Fernandes, Daniela and Welsch, Robin and Buschek, Daniel},
  journal={arXiv preprint arXiv:2509.11851},
  year={2025}
}

@article{draxler2024ai,
  title={The AI ghostwriter effect: When users do not perceive ownership of AI-generated text but self-declare as authors},
  author={Draxler, Fiona and Werner, Anna and Lehmann, Florian and Hoppe, Matthias and Schmidt, Albrecht and Buschek, Daniel and Welsch, Robin},
  journal={ACM Transactions on Computer-Human Interaction},
  volume={31},
  number={2},
  pages={1--40},
  year={2024},
  publisher={ACM New York, NY}
}

@article{savage2026ai,
  title={AI-Mediated Hiring and the Job Search of Blind and Low-Vision Individuals},
  author={Imteyaz, Kashif and Bart, Yakov and Das, Maitraye and Savage, Saiph and others},
  journal={arXiv preprint arXiv:2601.11884},
  year={2026}
}

@inproceedings{heyman2024supermind,
  title={Supermind ideator: How scaffolding human-AI collaboration can increase creativity},
  author={Heyman, Jennifer L and Rick, Steven R and Giacomelli, Gianni and Wen, Haoran and Laubacher, Robert and Taubenslag, Nancy and Knicker, Max and Jeddi, Younes and Ragupathy, Pranav and Curhan, Jared and others},
  booktitle={Proceedings of the ACM collective intelligence conference},
  pages={18--28},
  year={2024}
}

@inproceedings{birnholtz2012tracking,
  title={Tracking changes in collaborative writing: edits, visibility and group maintenance},
  author={Birnholtz, Jeremy and Ibara, Steven},
  booktitle={Proceedings of the ACM 2012 conference on computer supported cooperative work},
  pages={809--818},
  year={2012}
}

@article{liu2024survey,
  title={A survey of text watermarking in the era of large language models},
  author={Liu, Aiwei and Pan, Leyi and Lu, Yijian and Li, Jingjing and Hu, Xuming and Zhang, Xi and Wen, Lijie and King, Irwin and Xiong, Hui and Yu, Philip},
  journal={ACM Computing Surveys},
  volume={57},
  number={2},
  pages={1--36},
  year={2024},
  publisher={ACM New York, NY}
}

@incollection{weisz2025human,
  title={Human-Centered AI Design Principles for Generative AI},
  author={Weisz, Justin D and He, Jessica and Muller, Michael},
  booktitle={Handbook of Human-Centered Artificial Intelligence},
  pages={1--73},
  year={2025},
  publisher={Springer}
}

@inproceedings{mysore2025prototypical,
  title={Prototypical human-AI collaboration behaviors from LLM-Assisted writing in the wild},
  author={Mysore, Sheshera and Das, Debarati and Cao, Hancheng and Sarrafzadeh, Bahareh},
  booktitle={Proceedings of the 2025 Conference on Empirical Methods in Natural Language Processing},
  pages={16830--16857},
  year={2025}
}

@inproceedings{han2024teams,
  title={When teams embrace AI: human collaboration strategies in generative prompting in a creative design task},
  author={Han, Yuanning and Qiu, Ziyi and Cheng, Jiale and Lc, Ray},
  booktitle={Proceedings of the 2024 CHI Conference on Human Factors in Computing Systems},
  pages={1--14},
  year={2024}
}

@article{imteyaz2026upskilling,
  title={Upskilling with Generative AI: Practices and Challenges for Freelance Knowledge Workers},
  author={Imteyaz, Kashif and Lopez, Isabel and Rajpal, Nakul and Shin, Hunjun and Savage, Saiph},
  journal={arXiv preprint arXiv:2604.27231},
  year={2026}
}

@inproceedings{pachera2026co,
  title={Co-Data: Cultivating Effective Human-LLM Collaboration for Collaborative Data Processing},
  author={Pachera, Amedeo and Mauri, Andrea and Imteyaz, Kashif and Yang, Jie and Umuhoza, Eric and Bonifati, Angela and Lahav, Michal and Goyal, Nitesh},
  booktitle={Proceedings of the Extended Abstracts of the 2026 CHI Conference on Human Factors in Computing Systems},
  pages={1--7},
  year={2026}
}

@article{ulloa2025product,
  title={Product Manager Practices for Delegating Work to Generative AI:" Accountability must not be delegated to non-human actors"},
  author={Ulloa, Mara and Butler, Jenna L and Haniyur, Sankeerti and Miller, Courtney and Amos, Barrett and Sarkar, Advait and Storey, Margaret-Anne},
  journal={arXiv preprint arXiv:2510.02504},
  year={2025}
}

@inproceedings{he2025contributions,
  title={Which contributions deserve credit? Perceptions of attribution in human-AI co-creation},
  author={He, Jessica and Houde, Stephanie and Weisz, Justin D},
  booktitle={Proceedings of the 2025 CHI conference on human factors in computing systems},
  pages={1--18},
  year={2025}
}

@inproceedings{he2024ai,
  title={AI and the Future of Collaborative Work: Group Ideation with an LLM in a Virtual Canvas},
  author={He, Jessica and Houde, Stephanie and Gonzalez, Gabriel E and Silva Moran, Dar{\'\i}o Andr{\'e}s and Ross, Steven I and Muller, Michael and Weisz, Justin D},
  booktitle={Proceedings of the 3rd Annual Meeting of the Symposium on Human-Computer Interaction for Work},
  pages={1--14},
  year={2024}
}

@inproceedings{he2025exploring,
  title={Exploring Industry Practices and Perspectives on AI Attribution in Co-Creative Use Cases.},
  author={He, Jessica and Do, Hyo Jin},
  year={2025}
}

@inproceedings{hoque2024hallmark,
  title={The HaLLMark effect: Supporting provenance and transparent use of large language models in writing with interactive visualization},
  author={Hoque, Md Naimul and Mashiat, Tasfia and Ghai, Bhavya and Shelton, Cecilia D and Chevalier, Fanny and Kraus, Kari and Elmqvist, Niklas},
  booktitle={Proceedings of the 2024 CHI Conference on Human Factors in Computing Systems},
  pages={1--15},
  year={2024}
}

@article{savagegigsense,
  title={GigSense: An LLM-Infused Tool for Workers Collective Intelligence},
  author={Imteyaz, Kashif and Flores-Saviaga, Claudia and Savage, Saiph},
  journal={arXiv preprint arXiv:2405.02528},
  year={2024}
}

@inproceedings{saiph2024human,
  title={Human Computation, Equitable, and Innovative Future of Work AI Tools},
  author={Imteyaz, Kashif and Saviaga, Claudia Flores and Savage, Saiph},
  booktitle={Proceedings of the AAAI Conference on Human Computation and Crowdsourcing},
  volume={12},
  pages={155--156},
  year={2024}
}

@inproceedings{johnson2025augmenting,
  title={Augmenting Collaborative Problem-Solving: Exploring the Design and Use of GenAI for Groupwork},
  author={Johnson, Janet G and Rick, Steven R and Gr{\o}nb{\ae}k, Jens Emil and Wong, Emily and Yin, Ming and Nebeling, Michael and Klein, Mark and Ackerman, Mark S and Malone, Thomas},
  booktitle={Companion Publication of the 2025 Conference on Computer-Supported Cooperative Work and Social Computing},
  pages={168--173},
  year={2025}
}

@inproceedings{rifat2023many,
  title={Many Worlds of Ethics: Ethical Pluralism in CSCW},
  author={Rifat, Mohammad Rashidujjaman and Bhimdiwala, Ayesha and Bhattacharjee, Ananya and Batool, Amna and Das, Dipto and Mim, Nusrat Jahan and Safir, Abdullah Hasan and Sultana, Sharifa and Akter, Taslima and Smith, C Estelle and others},
  booktitle={Companion Publication of the 2023 Conference on Computer Supported Cooperative Work and Social Computing},
  pages={490--496},
  year={2023}
}

@inproceedings{rifat2024cohabitant,
  title={Cohabitant: The design, implementation, and evaluation of a virtual reality application for interfaith learning and empathy building},
  author={Rifat, Mohammad Rashidujjaman and Ayad, Reem and Asha, Ashratuz Zavin and Huang, Bingjian and Okman, Selin and Sabie, Dina and Ferdous, Hasan Shahid and Soden, Robert and Ahmed, Syed Ishtiaque},
  booktitle={Proceedings of the 2024 CHI Conference on Human Factors in Computing Systems},
  pages={1--19},
  year={2024}
}

@inproceedings{rifat2024data,
  title={Data, Annotation, and Meaning-Making: The Politics of Categorization in Annotating a Dataset of Faith-based Communal Violence},
  author={Rifat, Mohammad Rashidujjaman and Safir, Abdullah Hasan and Saha, Sourav and Junaed, Jahedul Alam and Saleki, Maryam and Amin, Mohammad Ruhul and Ahmed, Syed Ishtiaque},
  booktitle={Proceedings of the 2024 ACM Conference on Fairness, Accountability, and Transparency},
  pages={2148--2156},
  year={2024}
}

@inproceedings{ramesh2022platform,
  title={How platform-user power relations shape algorithmic accountability: A case study of instant loan platforms and financially stressed users in India},
  author={Ramesh, Divya and Kameswaran, Vaishnav and Wang, Ding and Sambasivan, Nithya},
  booktitle={Proceedings of the 2022 ACM conference on fairness, accountability, and transparency},
  pages={1917--1928},
  year={2022}
}

@inproceedings{ramesh2023ludification,
  title={Ludification as a lens for algorithmic management: A case study of gig-workers’ experiences of ambiguity in instacart work},
  author={Ramesh, Divya and Henning, Caitlin and Escher, Nel and Zhu, Haiyi and Lee, Min Kyung and Banovic, Nikola},
  booktitle={Proceedings of the 2023 ACM Designing Interactive Systems Conference},
  pages={638--651},
  year={2023}
}

\newpage

\appendix

\end{document}